\documentclass[a4paper,12pt]{article}
\usepackage{epsfig,float}
\usepackage{cite}
\usepackage[left=2cm,right=2cm]{geometry}
\usepackage{a4wide,graphicx}
\newcommand{\be}{\begin{equation}}
\newcommand{\ee}{\end{equation}}
\newcommand{\ba}{\begin{eqnarray}}
\newcommand{\ea}{\end{eqnarray}}

\begin{document}

\title{Dynamically generated resonances from the vector octet-baryon octet
interaction}

\author{E. Oset$^1$ and A. Ramos$^2$}
\maketitle

\begin{center}
$^1$ Departamento de F\'{\i}sica Te\'orica and IFIC,
Centro Mixto Universidad de Valencia-CSIC,
Institutos de Investigaci\'on de Paterna, Aptdo. 22085, 46071 Valencia, Spain\\
$^2$ Departament d'Estructura i
  Constituents de la Mat\`eria and Institut de Ci\`encies del Cosmos,
  Universitat de Barcelona, 08028 Barcelona, Spain\\
\end{center}

\date{}

 \begin{abstract}

 We study the interaction of vector mesons with the octet of stable baryons in
 the framework of the local hidden gauge formalism using a coupled channels unitary approach. We examine the scattering amplitudes and their poles, which
 can be associated to known $J^P=1/2^-,3/2^-$ baryon resonances, in some cases,
 or give predictions in other ones. The formalism employed produces doublets of degenerate
 $J^P=1/2^-,3/2^-$ states, a pattern which is observed experimentally in
 several cases. The findings of this work should also be useful to guide
 present experimental programs searching for new resonances, in particular
 in the strange sector where the current information is very poor.

\end{abstract}

\section{Introduction}

The use of chiral Lagrangians in combination with unitary techniques in coupled
channels of mesons and baryons has been a very fruitful scheme to study the 
nature of many hadron resonances. The analysis of meson
baryon scattering amplitudes shows poles in the second Riemann sheet which are
identified with existing baryon resonances. In this way the interaction of the octet of
pseudoscalar mesons with the octet of stable baryons has lead to $J^P=1/2^-$
resonances which fit quite well the spectrum of the known low lying resonances
with these quantum numbers
\cite{Kaiser:1995cy,weise,Kaiser:1996js,angels,ollerulf,carmina,carmenjuan,hyodo,Hyodo:2006kg}.
Similarly, the interaction
of the octet of pseudoscalar mesons with the decuplet of baryons also leads to
many resonances that can be identified with existing ones of $J^P=3/2^-$
\cite{kolodecu,sarkar}. Sometimes a new
resonance is predicted, as in the case of the $\Lambda(1405)$, where all the
chiral approaches find two poles close by, rather than one, a fact that finds
experimental support in the analyses of  Refs.~\cite{magas,sekihara}. The nature of the resonances
is admittedly more complex than just a molecule of pseudoscalar and baryon, 
but the success of this picture in reproducing many experimental
data on decay and production of the resonances provides support to claim very
large components of this character for the resonance wave function. In some cases
one can even reach the limits of the model and find observables that call for
extra components, even if small, but essential to explain some data. This can be
seen for instance in the radiative decay of the $\Lambda(1520)$ \cite{doring},
or in the helicity form factors of the $N^*(1535)$ \cite{jidomisha}. Another
promising approach is the one followed in \cite{hyodojido}, based on the
naturalness of the subtraction constants in the dispersion relations that hint
in some cases at the existence of non meson baryon components, particularly 
in the case of the
$N^*(1535)$. Another step forward in this direction has been the interpretation
of low lying $J^P=1/2^+$ as molecular states of two pseudoscalar mesons and one baryon
\cite{alberto,alberto2,kanchan,Jido:2008zz,KanadaEn'yo:2008wm}.

Much work has been done using pseudoscalar
mesons as building blocks, but the consideration of vectors instead of
pseudoscalars is only beginning to be exploited. In the baryon sector the
interaction of the $\rho$ $\Delta$ has been recently addressed in
\cite{vijande}, where three degenerate $N^*$ states around 1800 MeV and three degenerate
$\Delta$ states around $1900$ MeV, with $J^P=1/2^-, 3/2^-, 5/2^-$, are found.
This work has been recently extended to the SU(3) space of vectors and baryons
of the decuplet in \cite{souravbao}.  The
underlying theory for this study is the hidden gauge formalism
\cite{hidden1,hidden2,hidden3}, which deals with the interaction of vector mesons and
pseudoscalars in a way respecting chiral dynamics, providing the interaction of
pseudoscalars among themselves, with vector mesons, and vector mesons among
themselves. It also offers a perspective on the chiral Lagrangians as limiting
cases at low energies of vector exchange diagrams occurring in the theory.
In a more recent work, looking for poles in the $\pi N$ scattering amplitudes, 
the $\rho N$ channel is also included \cite{mishajuelich} and a resonance 
around 1700 MeV is dynamically generated, having the strongest coupling to 
this later channel.

 In the meson sector, the interaction of $\rho \rho$ within this formalism has
been addressed in \cite{raquel}, where it has been shown to lead  to the
dynamical generation of the $f_2(1270)$ and $f_0(1370)$  meson resonances, with
a branching ratio for the sensitive $\gamma \gamma$ decay channel in good
agreement with experiment \cite{junko}. This work has been extended to the
interaction of the SU(3) vector mesons in \cite{gengvec}, where  several
known resonances are also dynamically generated.

  In the present work we study the interaction of the octet of vector mesons
with the octet of stable baryons, using the unitary approach in coupled
channels. We shall see that the scattering amplitudes lead to poles in the
complex plane which can be associated to some well known resonances. Under the
approximation of neglecting the three momentum of the particles versus their
mass, we obtain degenerate states of $J^P=1/2^-,3/2^-$, a pattern which
seems to be followed qualitatively by the experimental spectrum, although in
some cases the spin partners have not been identified. A different approach to
account for the vector-baryon interaction is the one followed in \cite{su6}
where pseudoscalar mesons, vector mesons and baryons mix, advocating SU(6)
symmetry for the interaction. The approach leads to the same pseudoscalar-baryon
interaction than the hidden gauge approach, but to different results when it comes
to the interaction of the vector mesons with baryons. In particular, the spin
degeneracy predicted by the hidden gauge approach does not show up in the matrix
elements of the potential in the SU(6) scheme.

\section{Formalism for $VV$ interaction}

We follow the formalism of the hidden gauge interaction for vector mesons of
\cite{hidden1,hidden2,hidden3} (see also \cite{hidekoroca} for a practical set of Feynman rules).
The  Lagrangian involving the interaction of
vector mesons amongst themselves is given by
\begin{equation}
{\cal L}_{III}=-\frac{1}{4}\langle V_{\mu \nu}V^{\mu\nu}\rangle \ ,
\label{lVV}
\end{equation}
where the symbol $\langle \rangle$ stands for the trace in the $SU(3)$ space
and $V_{\mu\nu}$ is given by
\begin{equation}
V_{\mu\nu}=\partial_{\mu} V_\nu -\partial_\nu V_\mu -ig[V_\mu,V_\nu]\ ,
\label{Vmunu}
\end{equation}
where  $g$ is
\begin{equation}
g=\frac{M_V}{2f}\ ,
\label{g}
\end{equation}
with $f=93$ MeV the pion decay constant. With the value of $g$ of eq. (\ref{g})
one fulfills the KSFR rule \cite{KSFR} which is tied to
vector meson dominance \cite{sakurai}. The magnitude $V_\mu$ is the $SU(3)$
matrix of the vectors of the octet of the $\rho$
\begin{equation}
V_\mu=\left(
\begin{array}{ccc}
\frac{\rho^0}{\sqrt{2}}+\frac{\omega}{\sqrt{2}}&\rho^+& K^{*+}\\
\rho^-& -\frac{\rho^0}{\sqrt{2}}+\frac{\omega}{\sqrt{2}}&K^{*0}\\
K^{*-}& \bar{K}^{*0}&\phi\\
\end{array}
\right)_\mu \ .
\label{Vmu}
\end{equation}

The lagrangian ${\cal L}_{III}$ gives rise to a contact term coming from
$[V_\mu,V_\nu][V_\mu,V_\nu]$
\begin{equation}
{\cal L}^{(c)}_{III}=\frac{g^2}{2}\langle V_\mu V_\nu V^\mu V^\nu-V_\nu V_\mu
V^\mu V^\nu\rangle\ ,
\label{lcont}
\end{equation}
as well as to a three
vector vertex from
\begin{equation}
{\cal L}^{(3V)}_{III}=ig\langle (\partial_\mu V_\nu -\partial_\nu V_\mu) V^\mu V^\nu\rangle
\label{l3V}\ .
\end{equation}

It is convenient to rewrite the Lagrangian of eq. (\ref{l3V}) as
\begin{eqnarray}
{\cal L}^{(3V)}_{III}=ig\langle V^\nu\partial_\mu V_\nu V^\mu-\partial_\nu V_\mu
V^\mu V^\nu\rangle \nonumber\\
=ig\langle V^\mu\partial_\nu V_\mu V^\nu-\partial_\nu V_\mu
V^\mu V^\nu\rangle \nonumber\\
=ig\langle (V^\mu\partial_\nu V_\mu -\partial_\nu V_\mu
V^\mu) V^\nu\rangle
\label{l3Vsimp}\ .
\end{eqnarray}
In this case one finds an analogy to the coupling of vectors to
 pseudoscalars given in the same theory by
\begin{equation}
{\cal L}_{VPP}= -ig \langle [
P,\partial_{\nu}P]V^{\nu}\rangle \ ,
\label{lagrVpp}
\end{equation}
where $P$ is the SU(3) matrix of the pseudoscalar fields.

In a similar way, one obtains the Lagrangian for the coupling of vector mesons to
the baryon octet given by
\cite{Klingl:1997kf,Palomar:2002hk} \footnote{Correcting a misprint in
\cite{Klingl:1997kf}}
\begin{equation}
{\cal L}_{BBV} =
g\left( \langle \bar{B}\gamma_{\mu}[V^{\mu},B]\rangle + 
\langle \bar{B}\gamma_{\mu}B \rangle \langle V^{\mu}\rangle \right)
\label{lagr82}
\end{equation}
where $B$ is now the SU(3) matrix of the baryon octet
\begin{equation}
B =
\left(
\begin{array}{ccc}
\frac{1}{\sqrt{2}} \Sigma^0 + \frac{1}{\sqrt{6}} \Lambda &
\Sigma^+ & p \\
\Sigma^- & - \frac{1}{\sqrt{2}} \Sigma^0 + \frac{1}{\sqrt{6}} \Lambda & n \\
\Xi^- & \Xi^0 & - \frac{2}{\sqrt{6}} \Lambda
\end{array} \ .
\right)
\end{equation}

\begin{figure}[htb]
\begin{center}
\includegraphics[width=0.7\textwidth]{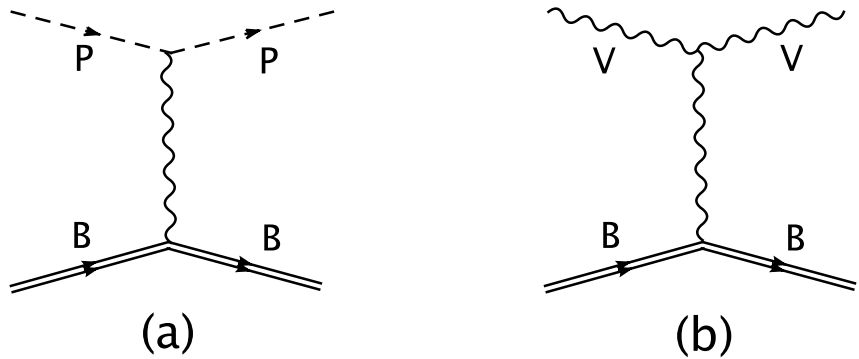}
\end{center}
\caption{Diagrams contributing to the pseudoscalar-baryon (a) or vector-
baryon (b) interaction via the exchange of a vector meson.}
\label{fig:feyn}
\end{figure}

With these ingredients we can construct the Feynman diagrams that lead to the $PB
\to PB$ and $VB \to VB$ interaction, by exchanging a vector meson between the
pseudoscalar or the vector meson and the baryon, as depicted in Fig.
\ref{fig:feyn}.

From the diagram of Fig. \ref{fig:feyn}(a), and under the low energy approximation of
neglecting $q^2/M_V^2$ in the propagator of the exchanged vector, where $q$ is the
momentum transfer, one obtains the
same amplitudes as obtained from the ordinary chiral Lagrangian for
pseudoscalar-baryon octet interaction \cite{Eck95,Be95}, namely the Weinberg-Tomozawa
terms.  One could anticipate some analogy between the vector-baryon
amplitudes with the pseudoscalar-baryon ones, given the
similarity of the Lagrangians in the way we have written them in 
Eqs.~(\ref{l3Vsimp}) and (\ref{lagrVpp}).
However, one also anticipates differences. Indeed, in the case of the 
pseudoscalar, Fig.~\ref{fig:feyn}(a), there is only one vector meson in the 
$PPV$ coupling and this must
necessarily be the one that is exchanged in the diagram.
However, in
the vector case, Fig.~\ref{fig:feyn}(b), the $VVV$ vertex involves three vector 
mesons and any of them can correspond in principle to the exchanged vector in 
the diagram. Nevertheless, one can proceed consequently
with neglecting $q^2/M_V^2$ as implicit in the chiral Lagrangians 
\cite{Eck95,Be95}, by also neglecting the three momentum of the
external vectors versus the vector mass. In this case, the
polarization vectors of the external vector mesons have only spatial components,
since the zeroth component is either zero for the transverse polarizations, or
negligible for the longitudinal one, ($k/\omega(k)$). Then, by examining the
Lagrangian of Eq.~(\ref{l3Vsimp})
one realizes that the field $V^\nu$ cannot
correspond to an external vector meson. Indeed, if this were the case, the $\nu$
index would be spatial and then the partial derivative $\partial_\nu$ would lead to
a three momentum of the vector mesons which are neglected in the approach. 
We can then conclude that the field $V^\nu$ corresponds to the exchanged vector
 and the analogy with the pseudoscalar and vector
interaction is then evident. Indeed, they are formally identical, by
substituting the octet of pseudoscalar fields by the octet of the vector fields,
with the additional factor $\vec{\epsilon}\,\vec{\epsilon }\,^\prime$ in the case of the
interaction of the vector mesons. Note that $\epsilon_\mu \epsilon^\mu$ becomes
$-\vec{\epsilon}\,\vec{\epsilon }\,^\prime$ and the signs of the Lagrangians also agree.

   A small amendment is in order in the case of vector mesons, which
   is due to the mixing of $\omega_8$ and the singlet of SU(3), $\omega_1$, to give the
   physical states of the $\omega$ and the $\phi$ mesons:
   \begin{eqnarray}
   \omega=\sqrt{\frac{2}{3}} \omega_1 + \frac{1}{\sqrt 3} \omega_8 \nonumber \\
   \phi=\frac{1}{\sqrt 3} \omega_1 - \sqrt{\frac{2}{3}} \omega_8
   \label{eq:omephi}
   \end{eqnarray}
   Given the structure of Eq.~(\ref{eq:omephi}), the singlet state  which is accounted for
   by the V matrix, $diag(\omega_1,\omega_1,\omega_1)/\sqrt3$, does not provide
   any contribution to Eq.~(\ref{l3Vsimp}), in which case all one must do is to take the
   matrix elements known for the $PB$ interaction and, wherever $P$ corresponds to the
   $\eta_8$, the amplitude should be multiplied by the factor $1/\sqrt 3$
   to get the corresponding $\omega $ contribution, and by $-\sqrt{2/3}$ to get the
   corresponding $\phi$ contribution. Upon the approximation consistent with
   neglecting the three momentum versus the mass of the particles (in this
   case the baryon), we can just take the $\gamma^0$ component of
   Eq. (\ref{lagr82})  and
   then the transition potential corresponding to the diagram of \ref{fig:feyn}(b) is
   given by
   \begin{equation}
V_{i j}= - C_{i j} \, \frac{1}{4 f^2} \, \left( k^0 + k^\prime{}^0\right)
~\vec{\epsilon}\,\vec{\epsilon }\,^\prime
\label{kernel}
\end{equation}
   where $k^0, k^\prime{}^0$ are the energies of the incoming and outgoing vector meson.

    The $C_{ij}$ coefficients of eq. (\ref{kernel}) can be obtained directly from
    \cite{angels,bennhold,inoue}
    with the simple rules given above for the $\omega$ and the $\phi$ mesons, and
    substituting $\pi$ by $\rho$ and $K$ by $K^*$ in the matrix elements. The
    coefficients are obtained both in the physical basis of states or in the
    isospin basis. Here we will directly study the interaction in isospin
    basis and in the appendix we collect the tables of the $C_{ij}$ 
    coefficients for different states of isospin, $I$, and strangeness, $S$.
     The tables immediately show the sectors where there is attraction 
     and therefore give 
      chances to find bound states or resonances. We can see that the
     cases with $(I,S)=(3/2,0)$, $(2,-1)$ and $(3/2,-2)$, the last two
     corresponding to exotic channels, are repulsive and do not
     produce poles
     in the scattering matrices.  However, the sectors
$(I,S)=(1/2,0)$, $(0,-1)$, $(1,-1)$ and $(1/2,-2)$ are attractive and we
expect to find bound states and resonances in these cases.

    The next step to construct the scattering matrix implies solving the
    coupled channels Bethe Salpeter equation in the on shell factorization approach of
    \cite{angels,ollerulf}
   \begin{equation}
T = [1 - V \, G]^{-1}\, V
\label{eq:Bethe}
\end{equation}
with $G$ being the loop function of a vector meson and a baryon which we calculate in
dimensional regularization using the formula of \cite{ollerulf}
\begin{eqnarray}
G_{l} &=& i 2 M_l \int \frac{d^4 q}{(2 \pi)^4} \,
\frac{1}{(P-q)^2 - M_l^2 + i \epsilon} \, \frac{1}{q^2 - m^2_l + i
\epsilon}  \nonumber \\ &=& \frac{2 M_l}{16 \pi^2} \left\{ a_l(\mu) + \ln
\frac{M_l^2}{\mu^2} + \frac{m_l^2-M_l^2 + s}{2s} \ln \frac{m_l^2}{M_l^2} +
\right. \nonumber \\ & &  \phantom{\frac{2 M}{16 \pi^2}} +
\frac{\bar{q}_l}{\sqrt{s}}
\left[
\ln(s-(M_l^2-m_l^2)+2\bar{q}_l\sqrt{s})+
\ln(s+(M_l^2-m_l^2)+2\bar{q}_l\sqrt{s}) \right. \nonumber  \\
& & \left. \phantom{\frac{2 M}{16 \pi^2} +
\frac{\bar{q}_l}{\sqrt{s}}}
\left. \hspace*{-0.3cm}- \ln(-s+(M_l^2-m_l^2)+2\bar{q}_l\sqrt{s})-
\ln(-s-(M_l^2-m_l^2)+2\bar{q}_l\sqrt{s}) \right]
\right\} \ ,
\label{eq:gpropdr}
\end{eqnarray}
with $\mu$ a regularization scale, which we take to be 630 MeV, and with a
natural value of the subtraction constant $a_l(\mu)$ of $-2$, 
as determined in \cite{ollerulf}.

 The iteration of diagrams implicit in the Bethe Salpeter equation in the case
 of the vector mesons has a subtlety with respect to the case of the
 pseudoscalars. The $\vec{\epsilon}\,\vec{\epsilon }\,^\prime$ term of the interaction
 forces the intermediate vector mesons in the loops to propagate with the
 spatial components in the loops.  We need to sum over the polarizations of the
 internal vector mesons which, because they are tied to the external ones
 through the $\vec{\epsilon}\,\vec{\epsilon }\,^\prime$ factor, provides
 \begin{equation}
\sum_{pol} \epsilon_i \epsilon_j =\delta_{ij} + \frac{q_i q_j}{M_V^2}
 \end{equation}
As shown in \cite{luisaxial}, the on shell factorization leads to a correction
coefficient in the $G$ function of $\vec{q}\,^{2}/3 M_V^2$ versus unity, which is negligibly small, and which we also neglect here in consonance with the approximations done. In this
case the factor $\vec{\epsilon}\,\vec{\epsilon }\,^\prime$, appearing in the potential $V$,
factorizes also in the $T$ matrix for the external vector mesons.

\section{Convolution due to the $\rho$ and $K^*$ mass distributions}

The formalism described above would provide results obtained using
fixed masses for the vector mesons and no width.
   The mass distributions of the $\rho$ and $K^*$ mesons are sufficiently extended
  to advise a more accurate calculation that takes this
 large width into account.
We follow the traditional method of convoluting the $G$ function with the
mass distributions of the  $\rho$ or $K^*$ mesons, as
is customarily made \cite{nagahiro}. One
  can prove that this convolution is equivalent to calculating the loop
  function with the dressed vector meson propagator written in terms of 
  its Lehmann representation, as is done in
  calculations of medium effects in the scattering matrices \cite{laura}.
The method amounts to replacing
the $G$ function by $\tilde{G}$ obtained as
\begin{eqnarray}
\tilde{G}(s)= \frac{1}{N}\int^{(m+2\Gamma_i)^2}_{(m-2\Gamma_i)^2}d\tilde{m}^2
\left(-\frac{1}{\pi}\right) 
{\rm Im}\,\frac{1}{\tilde{m}^2-m^2+{\rm i} \tilde{m} \Gamma(\tilde{m})}
& G(s,\tilde{m}^2,\tilde{M}^2_B)\ ,
\label{Gconvolution}
\end{eqnarray}
with
\begin{equation}
N=\int^{(m_\rho+2\Gamma_i)^2}_{(m_\rho-2\Gamma_i)^2}d\tilde{m}^2
\left(-\frac{1}{\pi}\right){\rm Im}\,\frac{1}{\tilde{m}^2-m^2_\rho+{\rm i} \tilde{m} \Gamma(\tilde{m})}
\label{Norm}
\end{equation}
being the normalization factor, and $\Gamma_i$ the decay width 
of the meson ($i=\rho,K^*$), which we take to be 149.4 MeV and 50.5 MeV for
the $\rho$ and  $K^*$ meson, respectively.
The energy dependent width $\Gamma(\tilde{m})$ for the $\rho$ meson,
obtained from its decay into two pions in $p$-wave,
is given by
\begin{equation}
\Gamma(\tilde{m})=\Gamma_\rho \frac{m_\rho^2}{\tilde{m}^2}
 \left(\frac{\tilde{m}^2-4m^2_\pi}{m^2_\rho-4m^2_\pi}\right)^{3/2}
 \theta(\tilde{m}-2m_\pi) \ .
\label{gamma}
\end{equation}
A similar expression gives the energy dependent width of the
$K^*$ meson from its decay into a $K$ meson and a pion.

We will see that, using fixed masses for the vector mesons, 
one finds bound states in the
$\rho N$ and $K^*N$ amplitudes, i.e. states having zero width. 
However, when 
$\tilde{G}$ is used in Eq.~(\ref{eq:Bethe}) and, therefore,
both the $\rho$ and $K^*$ vector mesons are taken with
their corresponding mass distribution, there is phase space for the decay of
each of these bound states into some of the mass components of the 
vector meson and 
the nucleon, thereby acquiring an appreciable width.

\section{Search for poles}

We search for poles in the scattering matrices in the second Riemann sheet, as
defined in previous works \cite{luisaxial}, basically changing $\bar{q}_l$ by to $-\bar{q}_l$ in the
analytical formula of the $G$ function, Eq. (\ref{eq:gpropdr}),
for channels where Re$(\sqrt s)$ is above the threshold of the corresponding
channel. When one has a mass distribution of the $\rho$ and $K^*$ mesons, and hence a fuzzy description of the threshold for some channels, one could take
different prescriptions for going to the optimal Riemann sheet that better
reflects the behavior of the amplitude in the real axis, which is where
the physical information is contained. 
The results are very similar in all cases, expect when one
has a resonance very close to threshold, where the convolution can distort the
shape of the amplitude and make even the pole disappear.  In view of that, 
for these cases the couplings are obtained from the amplitudes in the real axis 
as follows. Assuming these amplitudes to behave as
\begin{equation}
T_{ij}=\frac{g_i g_j}{\sqrt s -M_R + i \Gamma /2} \ ,
\end{equation}
where $M_R$ is the position of the maximum of $\mid T_{ii}\mid$, with $i$ being 
the channel to which the resonance couples more strongly, and $\Gamma$ its
width at half-maximum,
one then finds
\begin{equation}
\mid g_i \mid ^2= \frac{\Gamma}{2} \sqrt {|T_{ii}|^2} \ .
\label{couprealaxis}
\end{equation}
Up to a global phase, this expression allows one to determine the value of 
$g_i$, which we take to be real. The other couplings are then derived from
\begin{equation}
g_j = g_i \frac{T_{ij}(\sqrt{s}=M_R)}{T_{ii}(\sqrt{s}=M_R)} \ .
\label{couprealaxis2}
\end{equation}

This procedure to obtain the couplings from  $|T|^2$ in the real axis was used
in \cite{junko} where it was found that changes in the input parameters that lead to
moderate changes in the position and the width of the states affected the
couplings more smoothly.  
Having this in mind, we also calculate, as a check, the couplings in the
case of the particles without width from the residues of the amplitudes in the
first Riemann sheet of the complex plane. The couplings obtained
in this way are very similar to those obtained by means of 
Eqs.~(\ref{couprealaxis})
and (\ref{couprealaxis2}) employing
amplitudes in the real axis obtained with the convolution method. The differences
between the couplings obtained with the two methods are of the order of
10-20~\%.

\section{Results}

In this section we show our results obtained
in the attractive sectors
mentioned above. 

\begin{figure}[p]
\begin{center}
\includegraphics[width=0.55\textwidth]{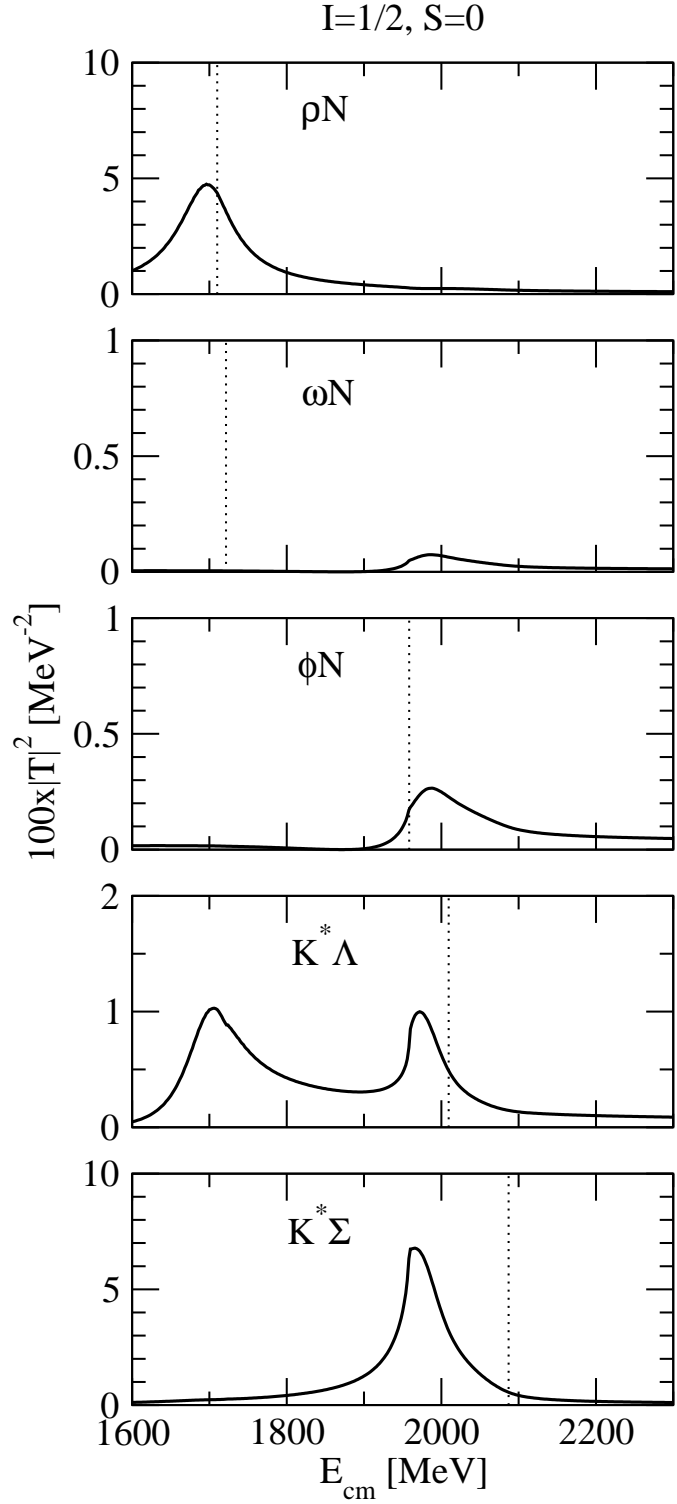}
\end{center}
\caption{$|T|^2$ for different channels for $I=1/2$ and strangeness $S=0$.
Channel thresholds are indicated by vertical dotted lines.}
\label{fig:s0}
\end{figure}

In Fig.~\ref{fig:s0} we show the results of $|T_{ii}|^2$ as a function of 
$\sqrt{s}$ 
for the different channels in the $(I,S)=(1/2,0)$ sector. We can see
a neat bump for  $\rho N \to \rho N$  around 1700
MeV, a few MeV below the  $\rho N$ threshold. The width of the peak is due to
the convolution of the $\rho$ mass. For a  $\rho$ of fixed mass and no width
one obtains a bound state with a small binding. The bump around 1700 MeV is also
seen in the $K^* \Lambda$ channel but is absent or barely visible in the
$\omega N$, $\phi N$ and $K^* \Sigma$ channels. On the other hand, with the
same strangeness and isospin, one finds a second peak around 1970 MeV, which is
clearly visible in the $K^* \Lambda$, $\omega N$, $\phi N$ and $K^* \Sigma$
channels but not visible in the $\rho N$ one. This special behavior has to do
with the coupling of the resonances to the different channels which will be
studied below.

  \begin{figure}[p]
\begin{center}
\includegraphics[width=0.80\textwidth]{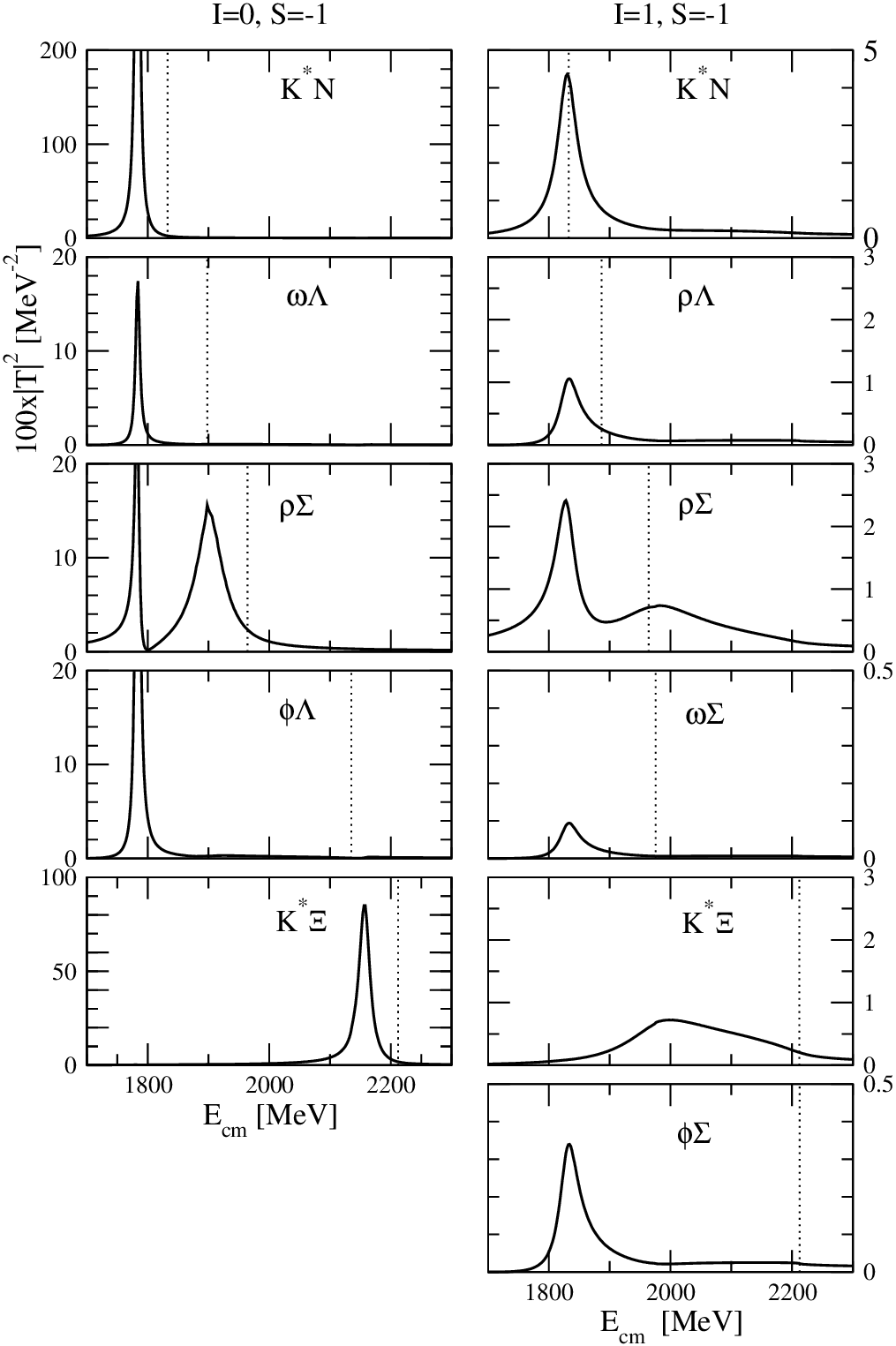}
\end{center}
\caption{$|T|^2$ for different channels for $I=0,1$ and strangeness $S=-1$.
Channel thresholds are indicated by vertical dotted lines.}
\label{fig:sm1}
\end{figure}

  \begin{figure}[p]
\begin{center}
\includegraphics[width=0.55\textwidth]{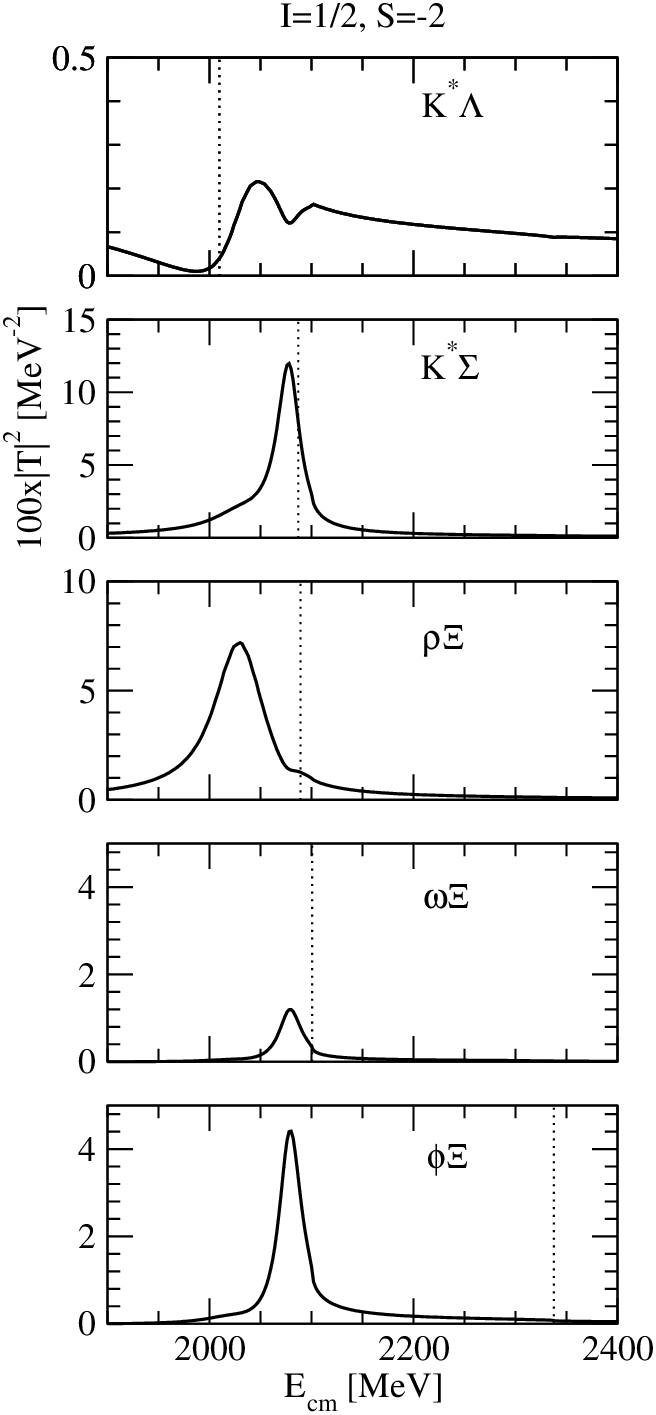}
\end{center}
\caption{$|T|^2$ for different channels for $I=1/2$ and strangeness $S=-2$.
Channel thresholds are indicated by vertical dotted lines.}
\label{fig:sm2}
\end{figure}

  In Fig.~\ref{fig:sm1} we show $|T_{ii}|^2$ for two cases, $(I,S)=(0,-1)$ and
  $(1,-1)$. For $I=0$ we see again a clear peak behavior of $|T_{ii}|^2$ around 1800
  MeV in the $\bar K^* N$ channel. The structure is also seen in all the other
  channels, except in the $K^* \Xi$ one. In the $\rho \Sigma$ channel one can
  see another structure around 1920 MeV, below the $\rho \Sigma$ threshold,
  which could be associated to a bound state of this system. Finally in the
  $K^* \Xi$ channel one finds another clear structure corresponding to a
  resonance around 2140 MeV, practically invisible in the other  channels,
  which could qualify as a  $K^* \Xi$ bound state.

   The $(I,S)=(1,-1)$ case shows a clear peak around 1830 MeV in the 
   $\bar K^*  N$ channel. This structure is also seen in other channels, but it could be
  the cusp due to the opening of the $\bar K^* N$ channel. We will come back to
  this when we make a search for the poles of the $T$ matrix.  In the $\rho
  \Sigma$ channel, and also visible in the $K^* \Xi$  channel, we see another
  less pronounced structure around 1990 MeV.

  In Fig.~\ref{fig:sm2} we display the sector $(I,S)=(1/2,-2)$. In this case we find a clear
  structure of resonant type around 2030 MeV which is visible in the $\rho \Xi$
  and $\bar K^* \Lambda$ channels. Another clear peak appears around 2080 MeV,  
  shortly below the $K^* \Sigma$ threshold, which is more pronounced in the $K^*\Sigma$ channel and
  appears as an interference
  minimum in the $\bar K^* \Lambda$ channel.

    Since the spin dependence only comes from the
 $\vec{\epsilon}\,\vec{\epsilon }\,^\prime$ factor and there is no dependence on the spin
  of the baryons, the interaction for vector-baryon states with $1/2^-$ and
  $3/2^-$ is the same and then one should associate each peak with the existence of two degenerate states. The spin degeneration also appears in some quarks models
 \cite{kirchbach}.

   The couplings of the resonances to the different channels,
    obtained from the residues at the poles are
   shown in Tables
   \ref{tab:S0I12} to \ref{tab:S2I12}.

\begin{table}[ht]
      \renewcommand{\arraystretch}{1.1}
     \setlength{\tabcolsep}{0.4cm}
\begin{center}
\begin{tabular}{c|cc|cc}\hline
$z_R$ & \multicolumn{2}{c|}{$1696^{(*)}$} &
\multicolumn{2}{c}{$1977 + {\rm i}53$} \\
\hline
 & $g_i$ & $\mid g_i\mid$ & $g_i$ & $\mid g_i\mid$ \\
\hline
$\rho N(1710)$ & $3.2 + {\rm i}0$ & 3.2 & $-0.3 -{\rm i} 0.5$ & 0.6  \\
$\omega N(1721) $ & $0.1 + {\rm i}0 $ & 0.1 & $-1.1 -{\rm i} 0.4$ & 1.2 \\
$\phi N(1958) $ & $ -0.2 + {\rm i}0 $ &  0.2 & $1.5 +{\rm i} 0.6$ & 1.7 \\
$K^* \Lambda(2010) $ & $ 2.3 + {\rm i}0 $ & 2.3 & $2.2 -{\rm i} 0.9$ & 2.3 \\
$K^* \Sigma(2087) $ & $ -0.6 + {\rm i}0 $ & 0.6 & $3.9 +{\rm i} 0.2$ & 3.9  \\
\hline
\end{tabular}
\caption{Pole position and coupling constants to various channels
 of the resonances found in the $I=1/2, S=0$ sector. ${}^{(*)}$ The properties of
 this state have been determined from the amplitudes in the real axis, as explained
 in the text.}
 \label{tab:S0I12}
\end{center}
\end{table}

\begin{table}[ht]
      \renewcommand{\arraystretch}{1.1}
     \setlength{\tabcolsep}{0.4cm}
\begin{center}
\begin{tabular}{c|cc|cc|cc}\hline
$z_R$ & \multicolumn{2}{c|}{$1784 + {\rm i}4$} &
\multicolumn{2}{c|}{$1906+{\rm i}70$} &
\multicolumn{2}{c}{$2158+{\rm i}13$} \\
\hline
 & $g_i$ & $\mid g_i\mid$ & $g_i$ & $\mid g_i\mid$ & $g_i$ & $\mid g_i\mid$ \\
\hline
${\bar K}^* N(1833)$ & $3.3 + {\rm i}0.07$ & 3.3 & $0.1 +{\rm i} 0.2$ & 0.3
& $0.2 +{\rm i} 0.3$ & 0.3 \\
$\omega \Lambda(1898)$ & $ 1.4 + {\rm i}0.03 $ &  1.4 & $0.4 +{\rm i} 0.2$ & 0.5
& $-0.3 -{\rm i} 0.2$ & 0.4 \\
$\rho \Sigma(1964)  $ & $-1.5  + {\rm i}0.03 $ & 1.5 & $3.1 +{\rm i} 0.7$ & 3.2
& $0.01 -{\rm i} 0.08$ & 0.08\\
$\phi \Lambda (2135)$ & $ -1.9 - {\rm i}0.04 $ & 1.9 & $-0.6 -{\rm i} 0.3$ & 0.6
& $0.5 +{\rm i} 0.3$ & 0.5  \\
$K^* \Xi (2212)$ & $  0.1 + {\rm i}0.003 $ & 0.1 & $0.3 +{\rm i} 0.1$ & 0.3
& $3.2 -{\rm i} 0.1$ & 3.2  \\
\hline
\end{tabular}
\caption{Pole position and coupling constants to various channels
 of the resonances found in the $I=0, S=-1$ sector.}
 \label{tab:S1I0}
\end{center}
\end{table}

\begin{table}[ht]
      \renewcommand{\arraystretch}{1.1}
     \setlength{\tabcolsep}{0.4cm}
\begin{center}
\begin{tabular}{c|cc|cc}\hline
$z_R$ & \multicolumn{2}{c}{$1830 + {\rm i}40^{(*)}$} &
\multicolumn{2}{c}{$1987+{\rm i}240^{(*)}$} \\
\hline
 & $g_i$ & $\mid g_i\mid$ & $g_i$ & $\mid g_i\mid$ \\
\hline
${\bar K}^* N(1833)$ & $2.1 + {\rm i}0$ & 2.1 & $-0.3-{\rm i}1.0 $ & 1.0   \\
$\rho \Lambda(1887) $ & $ -1.6 + {\rm i}0.2 $ &  1.6 & $-0.3 + {\rm i}0.9$ & 0.9 \\
$\rho \Sigma(1964) $ & $-1.6  + {\rm i}0.07 $ & 1.6 & $2.6 +{\rm i}0$ & 2.6  \\
$\omega\Sigma(1976) $ & $ -0.9 + {\rm i}0.1 $ & 0.9 & $-0.2  + {\rm i}0.5 $ & 0.5 \\
$K^* \Xi(2212) $ & $ 0.1 + {\rm i}0.06 $ & 0.1 & $2.1 - {\rm i}0.8$ & 2.3 \\
$\phi \Sigma(2213) $ & $  1.2 - {\rm i}0.2 $ & 1.2  & $0.2  -{\rm i}0.7 $ & 0.7 \\
\hline
\end{tabular}
\caption{Pole position and coupling constants to various channels
 of the resonances found in the $I=1, S=-1$ sector. ${}^{(*)}$ The properties of
 this state have been determined from the amplitudes in the real axis, as explained
 in the text.}
\label{tab:S1I1}
\end{center}
\end{table}

\begin{table}[ht]
      \renewcommand{\arraystretch}{1.1}
     \setlength{\tabcolsep}{0.4cm}
\begin{center}
\begin{tabular}{c|cc|cc}\hline
$z_R$ & \multicolumn{2}{c|}{$2039+ {\rm i}67$} &
\multicolumn{2}{c}{$2082 + {\rm i}31$} \\
\hline
 & $g_i$ & $\mid g_i\mid$ & $g_i$ & $\mid g_i\mid$ \\
\hline
${\bar K}^*\Lambda(2010) $ & $-0.7 - {\rm i}0.5 $ & 0.9 & $-0.1 -{\rm i} 0.3$ & 0.4 \\
${\bar K}^*\Sigma(2087)  $ & $ -0.9 - {\rm i}0.5 $ &  1.0 & $1.8 +{\rm i} 0.5$ &
1.9 \\
$\rho \Xi(2089)$   & $2.4 + {\rm i}0.7$ & 2.5 & $ 0.4 + {\rm i} 0.3$ & 0.5  \\
$\omega\Xi(2101)$   & $ 0.6 - {\rm i}0.08 $ & 0.6 & $1.1 +{\rm i} 0.3$ & 1.2 \\
$\phi\Xi(2038) $  & $ -0.8 + {\rm i}0.1 $ & 0.8 & $-1.6 -{\rm i} 0.4$ & 1.6  \\
\hline
\end{tabular}
\caption{Pole position and coupling constants to various channels
 of the resonances found in the $I=1/2, S=-2$ sector.}
 \label{tab:S2I12}
\end{center}
\end{table}

\newpage

\section{Comparison to data}

In table \ref{tab:pdg} we show a summary of the results obtained and the tentative
association to known states \cite{pdg}.

\begin{table}[ht]
      \renewcommand{\arraystretch}{1.5}
     \setlength{\tabcolsep}{0.2cm}
\begin{center}
\begin{tabular}{c|c|cc|ccccc}\hline\hline
$I,\,S$&\multicolumn{3}{c|}{Theory} & \multicolumn{5}{c}{PDG data}\\
\hline
    \vspace*{-0.3cm}
    & pole position    & \multicolumn{2}{c|}{real axis} &  &  & &  &  \\
    &   & mass & width &name & $J^P$ & status & mass & width \\
    \hline
$1/2,0$ & --- & 1696  & 92  & $N(1650)$ & $1/2^-$ & $\star\star\star\star$ & 1645-1670
& 145-185\\
  &      &       &     & $N(1700)$ & $3/2^-$ & $\star\star\star$ &
	1650-1750 & 50-150\\
       & $1977 + {\rm i} 53$  & 1972  & 64  & $N(2080)$ & $3/2^-$ & $\star\star$ & $\approx 2080$
& 180-450 \\	
   &     &       &     & $N(2090)$ & $1/2^-$ & $\star$ &
 $\approx 2090$ & 100-400 \\
 \hline
$0,-1$ & $1784 + {\rm i} 4$ & 1783  & 9  & $\Lambda(1690)$ & $3/2^-$ & $\star\star\star\star$ &
1685-1695 & 50-70 \\
  &       &       &    & $\Lambda(1800)$ & $1/2^-$ & $\star\star\star$ &
1720-1850 & 200-400 \\
       & $1907 + {\rm i} 70$ & 1900  & 54  & $\Lambda(2000)$ & $?^?$ & $\star$ & $\approx 2000$
& 73-240\\
       & $2158 + {\rm i} 13$ & 2158  & 23  &  &  &  & & \\
       \hline
$1,-1$ &  ---  & 1830  & 42  & $\Sigma(1750)$ & $1/2^-$ & $\star\star\star$ &
1730-1800 & 60-160 \\
  & ---    & 1987  & 240  & $\Sigma(1940)$ & $3/2^-$ & $\star\star\star$ & 1900-1950
& 150-300\\
   &     &       &   & $\Sigma(2000)$ & $1/2^-$ & $\star$ &
$\approx 2000$ & 100-450 \\\hline
$1/2,-2$ & $2039 + {\rm i} 67$ & 2039  & 64  & $\Xi(1950)$ & $?^?$ & $\star\star\star$ &
$1950\pm15$ & $60\pm 20$ \\
         & $2083 + {\rm i} 31 $ &  2077     & 29  &  $\Xi(2120)$ & $?^?$ & $\star$ &
$\approx 2120$ & 25  \\
 \hline\hline
    \end{tabular}
\caption{The properties of the 9 dynamically generated resonances and their possible PDG
counterparts.}
\label{tab:pdg}
\end{center}
\end{table}

  For the $(I,S)=(1/2,0)$ $N^*$  states there is the $N^*(1700)$ with
 $J^P=3/2^-$, which could correspond to the state we find with the same quantum
 numbers around the same energy. We also find in the PDG the  $N^*(1650)$, which
 could be the near degenerate spin parter of the $N^*(1700)$ that we predict in
 the theory. The difference of 50 MeV, is also the typical difference found in
 the $\Delta(1900)(1/2^-)$, $\Delta(1930)(5/2^-)$, $\Delta(1940)(3/2^-)$ states
 which are predicted as degenerate in the study of the interaction of vector
 mesons with the baryons of the decuplet in \cite{vijande,souravbao}.
 It is interesting to recall that in the study of Ref.~\cite{mishajuelich}, 
 done within the framework of the Juelich model
\cite{juelichmodel}, a pole is found around 1700 MeV, 
with the largest coupling to $\rho N$ states. 
The coupling found there is such that the strongest strength corresponds, 
by far, to the $L=0$, $J^P=3/2^-$ channel
\cite{mishaprivate}, which corresponds to the channel obtained in our approach.
However, the pole in Ref. \cite{mishajuelich} moves away 
from this position when the $N^*(1520)$ resonance is introduced as a 
 genuine resonance in their model, due to the mechanism of pole repulsion 
 discussed in Ref. \cite{Doring:2009bi}.

 Around 2000 MeV, where we find another $N^*$ resonance,
there are the states $N^*(2080)$ and $N^*(2090)$, with $J^P=3/2^-$ and
$J^P=1/2^-$ respectively, showing a good approximate spin degeneracy.

For the case $(I,S)=(0,-1)$ there is in the PDG one state, the $\Lambda(1800)$
with $J^P=1/2^-$, remarkably close to the energy were we find a $\Lambda$
state. The spin parter with $J^P=3/2^-$ is either absent in the PDG, or
corresponds to the $\Lambda(1690)$, although this implies a large breaking
of the expected degeneracy. The state obtained around 1900 MeV could
correspond to the $\Lambda(2000)$ cataloged in the PDG with unknown spin and parity.
On the other hand, one does not find in the PDG a resonance to associate with the
$\Lambda$ state predicted around 2150 MeV.

 The case of the $\Sigma $ states having $(I,S)=(1,-1)$ is rather interesting. The sate
that we find around 1830 MeV, could be associated to the  $\Sigma(1750)$
with $J^P=1/2^-$. More interesting seems to be the case of the state obtained around
1990 MeV that could be related to two PDG candidates, again
nearly degenerate, the $\Sigma(1940)$ and the $\Sigma(2000)$, with spin and
parity  $J^P=3/2^-$ and $J^P=1/2^-$ respectively.

  Finally, for the case of the cascade resonances, $(I,S)=(1/2,-2)$, we find 
  two states, one  around 2040 MeV and the other one around 2080 MeV. There are two cascade states in
  the PDG around this energy region with spin parity unknown, the
  $\Xi(1950)$ and the $\Xi(2120)$. The
  relatively small widths obtained in each case and the agreement with the
  experimental ones would be an extra feature to support the association of our states to
  these resonances. Although the experimental
  knowledge of this sector is relatively poor, a program is presently running at
  Jefferson Lab to improve on this situation \cite{Nefkens:2006bc}. 

    The agreement found in general is encouraging. One should stress that the
   measurements of the masses and widths in this energy region are not easy, as
   one can guess from the dispersion of the data obtained in different
   experiments.  The fact that some of the states predicted (essentially the
   spin partners) are not found in the PDG, should not be seen as a negative
   result of the theory, but as a motivation for the search of new resonances.
   The theory tells us the origin of these states, as coming from the vector baryon
   interaction. This gives us a new information and, although the
   larger part of the width may come from pseudoscalar baryon decay, it is in
   the vector baryon channels that experimental efforts should be made to
   eventually  find these states and confirm their vector-baryon nature.
   The devoted search of $\Xi$ resonances at
   Jefferson Lab  \cite{Nefkens:2006bc,Price:2004xm} should be most welcome in
   this context.

\section{Conclusions}

  We have studied the interaction of mesons in the vector octet of the $\rho$
  with baryons of the octet of the proton within the hidden gauge formalism of
  vector mesons, using a unitary framework in coupled channels. 
  
  We observe a rich
  structure in the vector-baryon scattering amplitudes which is associated
  to the presence of poles in
  the complex plane. This structure is clearly visible in the real axis as neat peaks of
  $|T|^2$ in different channels. We could associate many of the states predicted
  by the theory to known states in the PDG, thus providing a very different
  explanation for the nature of these states than the one given by quark models as simple $3q$
  states. One of the particular predictions of the theory is that, within the
  approximations done, one obtains degenerate pairs of particles in
  $J^P= 1/2^-,3/2^-$.  This behavior seems well reproduced by many of the
  existing data, but in some cases the spin partners do not show up in the PDG.
  The reasonable results produced by the hidden gauge approach in this case,
  as well as in other cases \cite{vijande,souravbao} should give a stimulus to
  search experimentally for the missing spin partners of the already observed
  states, as well as possible new ones.

\section*{Acknowledgments}

We would like to thank J. Garzon for checking the formulas of the paper. 
This work is partly supported by the EU contract No. MRTN-CT-2006-035482
(FLAVIAnet), by the contracts FIS2006-03438 FIS2008-01661 from MICINN
(Spain) and by the Ge\-ne\-ra\-li\-tat de Catalunya contract 2005SGR-00343. We
acknowledge the support of the European Community-Research Infrastructure
Integrating Activity ``Study of Strongly Interacting Matter'' (HadronPhysics2,
Grant Agreement n. 227431) under the Seventh Framework Programme of EU.

\appendix

\section{Coefficients of the $s$-wave tree level amplitudes}
\label{app:tables}

This Appendix gives the coefficients $C_{ij}^{IS}$ of the $s$-wave 
tree level vector-baryon  
amplitudes of Eq.~(\ref{kernel}) for the various $IS$ sectors studied in this
work.


\begin{table}[htbp]
 \renewcommand{\arraystretch}{1.2}
\centering

\caption{
 Coefficients $C_{ij}^{IS}$ for the sector $I=1/2$, $S=0$.
}
\vspace{0.5cm}
\begin{tabular}{l|ccccc}
 & $\rho N$ & $\omega N$ & $\phi N $  & $K^* \Lambda $ & $K^* \Sigma $\\
 \hline
$\rho N$ & 2 & 0 & 0 & $\frac{3}{2}$ & $-\frac{1}{2}$
  \\
$\omega N$ &  & 0 & 0 & $-\frac{3}{2}\frac{1}{\sqrt{3}}$ &
 $-\frac{3}{2}\frac{1}{\sqrt{3}}$ \\
$\phi N $ &  & & 0 & $-\frac{3}{2}
\left(-\sqrt{\frac{2}{3}}\right)$ &
 $-\frac{3}{2}
\left(-\sqrt{\frac{2}{3}}\right)$ \\
$K^* \Lambda $ &  & & & 0 & 0 \\
$K^* \Sigma $ &  & & & & 2  \\
\end{tabular}
\end{table}

\begin{table}[htbp]
 \renewcommand{\arraystretch}{1.2}
\centering

\caption{
 Coefficients $C_{ij}^{IS}$ for the sector $I=3/2$, $S=0$.
}
\vspace{0.5cm}
\begin{tabular}{l|cc}
 & $\rho N$ & $K^* \Sigma $ \\
 \hline
 $\rho N$ & $-1$ & $-1$ \\\
  $K^* \Sigma $ & & $-1$ \\
\end{tabular}
\end{table}

\begin{table}[htbp]
 \renewcommand{\arraystretch}{1.2}
\centering

\caption{
 Coefficients $C_{ij}^{IS}$ for the sector $I=0$, $S=-1$.
}
\vspace{0.5cm}
\begin{tabular}{l|ccccc}
& ${\bar K}^* N$ & 
$\omega \Lambda$ & 
$\rho \Sigma$ & 
$\phi \Lambda$ & 
$K^* \Xi$ \\
\hline
${\bar K}^* N$ & 3 & 
$\frac{3}{\sqrt{2}}\frac{1}{\sqrt{3}}$
& $-\sqrt{\frac{3}{2}}$ & 
$\frac{3}{\sqrt{2}}
\left(-\sqrt{\frac{2}{3}}\right)$
& 0 \\ 
$\omega \Lambda$ &  & 0 &  0 & 0 & 
$-\frac{3}{\sqrt{2}}\frac{1}{\sqrt{3}}$
\\
$\rho \Sigma$ & & & 4 & 0 & $\sqrt{\frac{3}{2}}$ \\
$\phi \Lambda$ & & & & 0 & 
$-\frac{3}{\sqrt{2}}
\left(-\sqrt{\frac{2}{3}}\right)$
\\
$K^* \Xi$ & & & & & 3 \\ 

\end{tabular}
\end{table}

\begin{table}[htbp]
 \renewcommand{\arraystretch}{1.2}
\centering

\caption{
 Coefficients $C_{ij}^{IS}$ for the sector $I=1$, $S=-1$.
}
\vspace{0.5cm}
\begin{tabular}{l|cccccc}
& ${\bar K}^* N$ &
$\rho \Lambda$ &
$\rho \Sigma$ &
$\omega \Sigma$ &  
$K^* \Xi$ &
$\phi \Sigma$\\
\hline
${\bar K}^* N$ & 1 & $-\sqrt{\frac{3}{2}}$ & 
$-1$ &
$-\sqrt{\frac{3}{2}}\frac{1}{\sqrt{3}}$  
& 0 
& $-\sqrt{\frac{3}{2}}
\left(-\sqrt{\frac{2}{3}}\right)$
\\
$\rho \Lambda$ & & 0   & 0  & 0  & 
$-\sqrt{\frac{3}{2}}$
& 0 \\
$\rho \Sigma$ & & & 2 & 0 & 1 & 0 \\
$\omega \Sigma$ & & & & 0 & 
$-\sqrt{\frac{3}{2}}\frac{1}{\sqrt{3}}$
& 0 \\ 
$K^* \Xi$ & & & & & 1  & 
$-\sqrt{\frac{3}{2}}
\left(-\sqrt{\frac{2}{3}}\right)$ \\
$\phi \Sigma$ & & & & & & 0 \\
\end{tabular}
\end{table}


\begin{table}[htbp]
 \renewcommand{\arraystretch}{1.2}
\centering
      
\caption{
 Coefficients $C_{ij}^{IS}$ for the sector $I=2$, $S=-1$.
}
\vspace{0.5cm}
\begin{tabular}{l|c}
 & $\rho \Sigma $ \\
 \hline
 $\rho \Sigma $ & $-2$\\
\end{tabular}
\end{table}   


\begin{table}[htbp]
 \renewcommand{\arraystretch}{1.2}
\centering

\caption{
 Coefficients $C_{ij}^{IS}$ for the sector $I=1/2$, $S=-2$.
}
\vspace{0.5cm}
\begin{tabular}{l|ccccc}
& ${\bar K}^* \Lambda$ &
${\bar K}^*  \Sigma$ &
$\rho \Xi$ &
$\omega \Xi$ &  
$\phi \Xi$\\
\hline
${\bar K}^* \Lambda$ & 0 & 0
& $-\frac{3}{2}$ 
& $-\frac{3}{2}\frac{1}{\sqrt{3}}$
& $-\frac{3}{2}\left(-\sqrt{\frac{2}{3}}\right)$
\\
${\bar K}^*  \Sigma$ & & 2 & 
$-\frac{1}{2}$ & $\frac{3}{2}\frac{1}{\sqrt{3}}$ &  
$\frac{3}{2}\left(-\sqrt{\frac{2}{3}}\right)$\\
$\rho \Xi$ & & & 2 
& 0
& 0\\
$\omega \Xi$ & & & & 0 & 0 \\  
$\phi \Xi$& & & & & 0 \\
\end{tabular}
\end{table}


\begin{table}[htbp]
 \renewcommand{\arraystretch}{1.2}
\centering
\caption{
 Coefficients $C_{ij}^{IS}$ for the sector $I=3/2$, $S=-2$.
}
\vspace{0.5cm}
\begin{tabular}{l|cc}
& ${\bar K}^* \Sigma$ &
$\rho \Xi$ \\
\hline
${\bar K}^*  \Sigma$ & $-1$ & $-1$ \\
$\rho \Xi$ & & $-1$ \\
\end{tabular}
\end{table}


\begin{table}[htbp]
 \renewcommand{\arraystretch}{1.2}
\centering

\caption{
 Coefficients $C_{ij}^{IS}$ for the sector $I=0$, $S=-3$.
}
\vspace{0.5cm}
\begin{tabular}{l|c}
 & ${\bar K}^* \Xi$  \\
 \hline
 ${\bar K}^* \Xi$ & 0\\
\end{tabular}
\end{table}

\begin{table}[h!]
 \renewcommand{\arraystretch}{1.2}
\centering
\caption{
 Coefficients $C_{ij}^{IS}$ for the sector $I=1$, $S=-3$.
}
\vspace{0.5cm}
\begin{tabular}{l|c}
 & ${\bar K}^* \Xi$  \\
 \hline
 ${\bar K}^* \Xi$ & $-2$\\
 \end{tabular}
\end{table}

\end{document}